\documentclass[conference]{IEEEtran}
\IEEEoverridecommandlockouts
\usepackage{amsfonts}
\usepackage{algorithmic}
\usepackage{algorithm}
\usepackage[dvips]{color}
\usepackage{tabu}
\usepackage{enumitem}
\usepackage{comment}
\usepackage{todonotes}
\usepackage{epsf}
\usepackage{times}
\RequirePackage{scrlfile}
\PreventPackageFromLoading{subfig}
\usepackage{bbold}
\usepackage{mathtools}
\usepackage{mathrsfs}
\usepackage{pdfpages}
\usepackage{epstopdf}
\usepackage{amsmath,epsfig,amssymb,amsthm,cite,url}
\usepackage{float}
\usepackage{graphicx}
\usepackage{textcomp}
\usepackage{geometry}
\usepackage{bbm}
\usepackage{xcolor}
\def\BibTeX{{\rm B\kern-.05em{\sc i\kern-.025em b}\kern-.08em
    T\kern-.1667em\lower.7ex\hbox{E}\kern-.125emX}}
\begin{document}
\title{\huge{Variational Autoencoders for Reliability Optimization in Multi-Access Edge Computing Networks}}
\DeclareRobustCommand*{\IEEEauthorrefmark}[1]{%
  \raisebox{0pt}[0pt][0pt]{\textsuperscript{\footnotesize #1}}%
}

\author{\IEEEauthorblockN{Arian Ahmadi,
Omid Semiari,
Mehdi Bennis, and
Mérouane Debbah
\thanks{This research was supported by the U.S. National Science Foundation under Grants CNS 1941348 and CNS 2008646.}}
\vspace{-1cm}
\thanks{A. Ahmadi and O. Semiari are with the Department  of Electrical and Computer Engineering, University of Colorado, Colorado Springs, CO, 80918 USA (Email: aahmadi@uccs.edu; osemiari@uccs.edu).
}
\thanks{M. Bennis is with the Department of Communications Engineering, University of Oulu, FI-90014, Oulu, Finland (Email: mehdi.bennis@oulu.fi).}
\thanks{M. Debbah is with the Technology Innovation Institute, (email: merouane.debbah@tii.ae) and also with the Mohamed Bin Zayed University of Artificial Intelligence, 9639 Masdar City, Abu Dhabi, United Arab Emirates.}
} 
\maketitle
\begin{abstract}
Multi-access edge  computing (MEC) is viewed as an integral part of future wireless networks to support new applications with stringent service reliability and latency requirements. However, guaranteeing ultra-reliable and low-latency MEC (URLL MEC) is very challenging due to uncertainties of wireless links, limited communications and computing resources, as well as dynamic network traffic. Enabling URLL MEC mandates taking into account the statistics of the end-to-end (E2E) latency and reliability across the wireless and edge computing systems. In this paper, a novel framework is proposed to optimize the reliability of MEC networks by considering the distribution of E2E service delay, encompassing over-the-air transmission and edge computing latency. The proposed framework builds on \emph{correlated} variational autoencoders (VAEs) to estimate the full distribution of the E2E service delay. Using this result, a new optimization problem based on risk theory is formulated to maximize the network reliability by minimizing the Conditional Value at Risk (CVaR) as a risk measure of the E2E service delay. To solve this problem, a new algorithm is developed to efficiently allocate users' processing tasks to edge computing servers across the MEC network, while considering the statistics of the E2E service delay learned by VAEs. The simulation results show that the proposed scheme outperforms several baselines that do not account for the risk analyses or statistics of the E2E service delay. 
\end{abstract}

\section{Introduction}
The sixth-generation (6G) wireless cellular network must support a broad range of new applications with ultra-high reliability and low latency service requirements \cite{semiari2019integrated}. Among these emerging applications include factory automation, connected and autonomous vehicles (CAVs), and extended reality (XR) with strict quality-of-service (QoS) requirements on the end-to-end (E2E) latency (e.g., 1 ms) and reliability (e.g., $10^{-8}$ packet loss probability) \cite{3GPP}. To meet such stringent service requirements, multi-access edge computing (MEC) is an attractive solution to significantly reduce service latency by enabling base stations (BSs) to process computing tasks (e.g., XR rendering) for user equipment (UE) directly within the radio access network (RAN) without relying on remote cloud servers \cite{pham2020survey, mao2017survey}.

While promising, delivering a guaranteed performance over MEC is challenging due to random wireless channel variations, stochastic task arrival, as well as heterogeneity of edge computing servers and computing tasks. Furthermore, given the resource limitations of wireless networks and edge servers (e.g., in terms of bandwidth or computing power), the MEC network can easily get overloaded, leading to computing task drops and poor QoS. Hence, it is imperative to develop novel solutions for optimizing the allocation of edge computing resources to UEs' processing tasks within the MEC, while taking into account the statistics of E2E reliability and latency across communications and computing systems.

Recently, a large body of work \cite{mao2017survey, jiang2019toward, pham2020survey, kovalenko2019robust, wu2018multi} has focused on ultra-reliable and low-latency MEC (URLL MEC). Most of these existing works do not rely on statistical analyses of E2E reliability and latency since deriving full distribution of these metrics via traditional model-based approaches is very challenging. For URLL MEC, in particular, there is a need to account for joint wireless and edge computing  constraints and conventional model-based approaches are not tractable in such complex network scenarios. As an alternative to model-based methods, model-free approaches based on machine learning (ML) along with computing capabilities of UEs and BSs can provide new opportunities for enabling URLL MEC.

In this regard, the body of work in \cite{jiang2016machine, ye2019deep, li2018deep, kasgari2020experienced} presents several new schemes based on deep neural networks (DNNs) and deep reinforcement learning (DRL) to optimize the network performance for URLL MEC applications. The authors in \cite{ye2019deep} propose a decentralized resource allocation technique using DRL for vehicular communications. In \cite{li2018deep}, the authors present a DRL-based resource allocation scheme for URLL MEC. However, most of existing works assume that explicit traffic and queue models are available to the resource management system. In \cite{kasgari2020experienced}, the concept of experienced deep reinforcement learning is proposed via generative adversarial networks (GANs) in order to provide model-free resource allocation for low latency services. However, existing model-free methods (e.g., in \cite{kasgari2020experienced}) do not account for the distribution of E2E reliability or latency or statistical correlation between computing and communication delays when designing task assignment in MEC networks. In contrast with the prior art, we develop a new ML-based approach for statistical optimization of E2E reliability in MEC networks.

The main contribution of this paper is a novel framework to optimize the reliability of MEC networks while considering the distribution of the \emph{E2E service delay}, comprising the over-the-air transmission and edge computing latencies. To derive the distribution of the E2E service delay, the proposed method builds on correlated variational autoencoders (VAEs) which are powerful deep generative models for estimating the probability density function (PDF) of latent
variables that represent the feature space of the training data. This is achieved by modeling the transmission and computing delays as two correlated Gaussian random variables (within a two-dimensional latent space). Then, using the derived distribution for the E2E service delay, we define a reliability metric for the MEC network and formulate a new optimization problem based on risk theory that aims to maximize the network reliability by efficiently allocating the edge computing resources to UEs' processing tasks across the RAN, while accounting for the constraints of the wireless network and edge computing servers.

In particular, the proposed optimization problem uses the concept of Conditional Value at Risk (CVaR) to capture the tail of the system's E2E service delay distribution learned by the correlated VAE. Given that the proposed problem is a mixed integer non-linear programming (MINLP) and difficult to solve, a new algorithm is developed to jointly solve the task allocation problem and computing resource allocation across the MEC. Comprehensive simulations are performed based on both synthetic and real data for a network of CAVs as a case study. The results show that the proposed correlated VAE scheme can effectively capture the correlation between over-the-air transmission and computing latencies resulting from an imbalanced load distribution in MEC. In addition, the results show that the proposed approach can guarantee 30 ms E2E service delay with a high probability of 0.9999 compared to two other baselines that can only satisfy this E2E latency requirement with probabilities less than 0.92. 

The rest of the paper is organized as follows. Section II presents the system model. Section III presents the proposed approach for deriving the distribution of the service delay. Section IV describes the problem formulation and the proposed solution. Simulation results are provided in Section V and conclusions are presented in Section VI.
\vspace{-0.15cm}
\section{System Model}
\label{System Model}
Consider an MEC network consisting of a set $\mathcal{M}$ of $M$ UEs and a set $\mathcal{N}$ of $N$ BSs. This network allows UEs to submit their computing tasks along with the associated data (e.g., camera and LiDAR outputs for rendering a high-definition map for CAVs) to a nearby BS. Each BS collects the requests from UEs in the uplink, uses its edge computing server to process the tasks, and sends the results back to its assigned UEs over downlink transmissions. Therefore, the \emph{E2E service delay} for a UE $m \in \mathcal{M}$ served by a BS $n \in \mathcal{N}$ can be defined as \cite{tareq2018ultra}
\begin{equation}\label{E2E latency}
\tau(m, n)=\left[\tau_{t}(m, n)\right] T+\tau_{p}(m, n),\vspace{-0.15cm}
\end{equation}
where $T$ is the duration of one transmission time interval (TTI) and $\tau_{t}(m, n)$ represents the total over-the-air transmission latency over uplink and downlink, in terms of the number of TTIs. In addition, $\tau_{p}(m, n)$ is the computational latency at the edge server of a BS $n$ to process the requested task by UE $m$. Using (\ref{E2E latency}), we can deﬁne the reliability of the link between BS $n$ and UE $m$ within the MEC network as 
\begin{equation}\label{R}
\operatorname{F}_{\tau(m,n)}(\tau_{\text{th}})=\operatorname{P}(\tau(m,n)<\tau_{\text{th}})  ,\vspace{-0.15cm}
\end{equation}
where $\tau_{\text{th}}$ is a delay threshold that is determined based on the QoS requirement. In fact, (\ref{R}) implies that the reliability depends on the cumulative distribution function (CDF) of the E2E service delay $\tau$. Next, we explain the transmission and computing latencies in details.

\subsection{Over-the-Air Transmission Latency}\label{Transmission}
The overall transmission latency consists of the downlink and uplink transmission latencies as follow: 
\begin{equation}\label{transmission latency}
\tau_{t}(m, n)=\tau_{d}(m, n)+\tau_{u}(m, n),
\end{equation}
where $\tau_{u}(m, n)$ represents the uplink transmission latency for UE $m$ to send its computing task request along with any associated data to BS $n$. Due to the wireless fading channel, some packets may not be decoded successfully at the BS, hence, retransmission is needed. With this in mind, the uplink transmission latency is given by: 
\begin{equation}\label{uplink latency}
\tau_{u}(m, n)= \sum_{ j =1}^{J}\left\lceil\frac{I_{u}}{R_{u j}(m, n) T}\right\rceil,\vspace{-0.15cm}
\end{equation}
where $\lceil.\rceil$ is the ceiling function, $J-1$ is the number of retransmissions, and $I_{u}$ is the uplink packet size (in bits). In (\ref{uplink latency}), $R_{uj}(m,n)$ represents the uplink data rate for the $j$-th transmission of a packet during one TTI and is given by:
\begin{equation}
R_{u j}(m, n)=w \log _{2}\left(1+\gamma_{u j}(m, n)\right),\vspace{-0.15cm}
\end{equation}
where $w$ is the channel bandwidth. Moreover, $\gamma_{u j}(m, n)$ is the uplink signal-to-interference-plus-noise ratio (SINR) and is given by
\begin{equation}\label{SINR_uplink}
\gamma_{u j}\left(m, n\right)=\frac{G_{m} G_{n} P_{m} h_{m n j} L_{m n}}{\sum_{m^{\prime} \neq m} P_{m^{\prime} n}+\sigma_{n}^{2}},\vspace{-0.15cm}
\end{equation}
where $P_m$, $P_{m^{\prime} n}$, and $\sigma_{n}^{2}$ denote, respectively, the transmit power of UE $m$, the received power from an interfering UE $m^{\prime}$, and the noise power. $G_m$ and $G_n$ are the antenna gains for UE $m$ and BS $n$, respectively. In addition, $h_{mnj}$, and $L_{mn}$ represent, respectively, the Rayleigh fading channel gain for the $j$-th transmission and path loss of the uplink between UE $m$ and BS $n$. The channel gain $h_{mnj}$ is considered flat-fading over the bandwidth $w$ and constant during the transmission of one packet.
Similar to (\ref{uplink latency}), the downlink transmission latency, $\tau_{d}(m, n)$, for sending the processed results from BS $n$ to UE $m$ will be
\begin{equation}\label{downlink latency}
\tau_{d}(m, n)=\sum_{i =1}^{I}\left\lceil\frac{ I_{d}}{R_{d i}(m, n) T}\right\rceil,\vspace{-0.15cm}
\end{equation}
where $I_{d}$ is the downlink packet size in bits and $I-1$ is the number of retransmissions. In (\ref{downlink latency}), $R_{di}(m,n)$ represents the downlink data rate for the $i$-th transmission of a packet and is given by:
\begin{equation}
R_{d i}(m, n)=w \log _{2}\left(1+\gamma_{d i}(m, n)\right).\vspace{-0.15cm}
\end{equation}
In (8), $\gamma_{d i}(m, n)$ is the downlink SINR which is given by
\begin{equation}\label{SINR_downlink}
\gamma_{d i}(m, n)=\frac{G_{m} G_{n} P_{n} h_{m n i} L_{m n}}{\sum_{n^{\prime} \neq n} P_{m n^{\prime}}+\sigma_{n}^{2}},\vspace{-0.15cm}
\end{equation} 
where $P_n$ and $P_{m n^{\prime}}$ denote, respectively, the transmit power of BS $n$ and the received power from interfering BS $n^{\prime}$ at the UE $m$’s receiver.

\subsection{Computing Latency at Edge Computing Servers }\label{Computational}
Computing latency refers to the time needed for executing a task at an edge computing server within the MEC network. The execution time of a task depends on the data to be processed and the tasks submitted, therefore, it can be modeled as a random variable \cite{tareq2018ultra}. For instance, the execution time for performing the object detection highly depends on the quality or level of details in the captured images, as well as the type (GPU vs. CPU) and processing resources (e.g., processing bandwidth) of the edge server.

Let $f_{max}$ denote the maximum computing-cycle frequency of edge processors for each BS, and $f_m$ as the computing-cycle frequency (i.e., the number of computing-cycles required to process
one bit of data) allocated to the task requested by UE $m$. Therefore, the total cycles needed for processing the submitted task by UE $m$ will be $c_m = f_m I_{u} $. In addition, we define $f(m, n)$ as the computing frequency allocated to UE $m$ by BS $n$. Therefore, the Computing latency in (\ref{E2E latency}) can be calculated as $\tau_{p}(m, n) = \frac{c_m}{f(m, n)}$.

Given the proposed model, deriving the distribution of the E2E service delay is very challenging due to the stochastic effects of the wireless fading channel, noise, random queuing delays at the transmitters and edge computing servers, packet retransmissions, along with heterogeneity of processing tasks and edge computing resources. To address this challenge, in the next section,  we will adopt a new approach based on VAEs to derive the distribution of the E2E service delay. While the proposed models in Section II are not used directly to derive the distribution of E2E service delay, they will be used to generate the data for training the VAE model.

\section{Correlated VAE for Learning the Distribution of the Service Delay}
Next, we first review the standard VAE. Then, we present the adopted correlated VAE model that will be used to capture the correlation between the transmission and computing latencies.
\vspace{-1em}
\subsection{Standard VAE model}
The standard VAE model consists of a DNN that performs as an encoder followed by a second DNN that serves as a decoder \cite{kingma2013auto}. The encoder of the VAE takes $\boldsymbol{x}=\left[{x}_{1}, \ldots, {x}_{D}\right] \subseteq \mathbbm{R}^{D}$ as input and encodes the data into a latent representation $\boldsymbol{z}=\left[{z}_{1}, \ldots, {z}_{d}\right] \subseteq \mathbbm{R}^{d}$. Using the latent representation, the decoder aims to generate a reconstruction data that is statistically similar to the input data. In the MEC network, BSs can train VAEs to learn the distribution of the service delay by considering the transmission and computing latencies as the two variables of the latent space (i.e., $\boldsymbol{z}=[\tau_{t},\tau_{p}]$).

More specifically, the VAE generates the latent representation $\boldsymbol{z}$ by drawing $L$ independent samples ${z}_{l} \stackrel{i . i . d}{\sim} p_{\theta}({z}_{l})$, $l \in \{1, 2, \cdots, L\}$, from the prior distribution $p_{\theta}$ which usually is a standard Gaussian distribution. Then, the decoder generates $L$ independent data points ${x}_{l} \sim p_{\boldsymbol{\theta}}\left({x}_{l} \mid {z}_{l}\right)$ from the model conditional distribution $p_{\boldsymbol{\theta}}$. The objective of the VAE is to maximize the marginal likelihood $p_{\boldsymbol{\theta}}(\boldsymbol{x})$, where $ \boldsymbol{\theta}$ is a vector of all parameters in the decoder $p_{\boldsymbol{\theta}}(\boldsymbol{x}|\boldsymbol{z})$. The marginal likelihood is intractable \cite{kingma2013auto}, and is approximated by the evidence lower bound (ELBO) defined as\vspace{-0.5em}
\begin{equation}\label{ELBO}  \mathcal{L}(\boldsymbol{\theta}, {\phi})=\langle\log p_{\boldsymbol{\theta}}(\boldsymbol{x}\vert \boldsymbol{z})\rangle_{q_{{\phi}(\boldsymbol{z}\vert \boldsymbol{x})}}-\mathrm{KL}[q_{{\phi}}(\boldsymbol{z}\vert \boldsymbol{x})\Vert p_{\boldsymbol{\theta}}(\boldsymbol{z})], \!\!\end{equation}
which satisfies $\mathcal{L}(\boldsymbol{\theta}, {\phi}) \leq \log p_{\boldsymbol{\theta}}(\boldsymbol{x})$. In (\ref{ELBO}), $q_{{\phi}}(\boldsymbol{z}|\boldsymbol{x})$ is the encoder network parameterized by ${\phi}$, KL$[\cdot \| \cdot]$ denotes Kullback-Leibler (KL) divergence, and $\langle\cdot\rangle_{p(\cdot)}$ is the expectation over a distribution $p(\cdot)$.

We use a finite set of $L$ samples of $\boldsymbol{z}$ to approximate the expectation in (\ref{ELBO}) as
\begin{equation}\label{loss}
\!\mathcal{L}(\boldsymbol{\theta},\! {\phi}) \simeq\frac{1}{L}\!\sum_{l}[\log p_{\boldsymbol{\theta}}(\boldsymbol{x}_{l}\vert \boldsymbol{z}_{l})-\mathrm{KL}[q_{{\phi}}(\boldsymbol{z}_{l}\vert \boldsymbol{x}_{l})\Vert p_{\boldsymbol \theta}(\boldsymbol{z}_{l})]].  \vspace{-0.15cm}
\end{equation} 
The VAE loss function is the sum of two loss terms – a reconstruction loss, and a latent loss. The former is used to fit the reconstructed vector to the original vector, whereas the latter pushes the variational distribution $q_{{\phi}}(\boldsymbol{z}\vert \boldsymbol{x})$ towards the prior distribution $p_{\theta}\left(\boldsymbol{z}\right)$. However, sampling $\boldsymbol{z}$ from $q_{{\phi}}(\boldsymbol{z}\vert \boldsymbol{x})$ is a non-differentiable operation. The work-around this issue is the “reparametrization trick” by creating randomness from a fixed distribution $\boldsymbol{\epsilon} \sim \mathcal{N}(\boldsymbol{\mu}_{prior},\boldsymbol{\sigma}_{prior}\mathbf{I_d})$. Then, the samples of the appropriate distribution can be generated by computing $\mathbf{z}_{l}=\boldsymbol{\mu}_{l}+\tilde{\mathbf{C}}_{l} \boldsymbol{\epsilon}$ where $\boldsymbol{\mu}_{l}$ is the mean vector of the posterior learned for the $l_{th}$ sample and $\tilde{\mathbf{C}}_{l}$ is the Choleskiy decomposition of the corresponding covariance matrix $\mathbf{C}_l$. While the prior distribution is chosen as $p_{\boldsymbol{\theta}}(\boldsymbol{z})= \mathcal{N}(\boldsymbol{\mu}_{prior},\boldsymbol{\sigma}_{prior}\mathbf{I_d})$, the standard choice widely adopted for the sample-wise approximate posterior is to set $\mathbf{C}_{l}^{(\mathbf{s})}=\operatorname{diag}\left(\mathbf{s}_l\right)$. In other words, $q_{\phi}\left(\mathbf{z}_{l} \mid \mathbf{x}_{l}\right)=\mathcal{N}\left(\boldsymbol{\mu}_{l}, \operatorname{diag}\left(\mathbf{s}_{l}\right)\right)$ hence, the  latent loss in (\ref{loss}) can be written as
\begin{equation}\label{VAELOSSFUNCTION}
\begin{aligned}
&\mathrm{KL}[q_{{\phi}}(\boldsymbol{z}_{l}\vert \boldsymbol{x}_{l})\Vert p_{\boldsymbol \theta}(\boldsymbol{z}_{l})]=\\&\frac{1}{2}\left[\frac{\mathbf{s}^2_{l}}{\boldsymbol{\sigma}^2_{prior}} +\frac{(\boldsymbol{\mu}_l-\boldsymbol{\mu}_{prior})^{2}}{\boldsymbol{\sigma}^2_{prior}}\right.\left.-d- \log (\frac{\mathbf{s}_{l}}{\boldsymbol{\sigma}_{prior}})\right].
\end{aligned}\vspace{-0.2cm}
\end{equation}

Since the variational distribution $q_{{\phi}}(\boldsymbol{z}\vert \boldsymbol{x})$ factorizes over input data points and the prior distribution is i.i.d. Gaussian, the KL-divergence in the ELBO is a sum over the per-data-point KL-divergence terms, which means that we disregard any correlation within dimensions of the latent space. In our problem, this will be a limiting aspect as the transmission and computing latencies can be correlated. For example, if a BS is overloaded (i.e., many UEs are assigned to one BS), both wireless and computing resources will be distributed among a large set of UEs, which can lead to an increase in both transmission and computing delays. 
\vspace{-0.15cm}
\subsection{Correlated VAE model}\label{correlated VAE}
Since the standard VAE does not allow any correlation between the dimensions of the approximate posterior, in this section, we implement a new class of VAEs that uses the first-order autoregressive Gaussian, $q_{\phi}\left(\mathbf{z}_{l} \mid \mathbf{x}_{l}\right)=\mathcal{N}\left(\boldsymbol{\mu}_{l}, \mathbf{C}_{(\rho,s)} \right)$, instead of the standard Gaussian with diagonal covariance. First-order autoregressive covariance, $\mathbf{C}_{(\rho,s)}$, is characterized by a scaling factor $s$ and a scalar $\rho$ to control the level of correlation and is defined as
\begin{equation}
\mathbf{C}_{(\rho,s)} = 
s\begin{bmatrix}
1 & \rho & \rho^2 & \ldots{} & \rho^{d-1}\\
\rho & 1 & \rho & \ldots{} & \rho^{d-2}\\
\rho^2 & \rho & 1 & \ldots{} & \rho^{d-3}\\
\vdotswithin{\ldots} &  \vdotswithin{\ldots} &  \vdotswithin{\ldots} & \ldots{} &  \vdotswithin{\ldots}\\
\rho^{d-1} & \ldots{} & \rho^2 & \rho & 1
\end{bmatrix},\vspace{-0.15cm}
\end{equation}
where $s$ is a positive scalar and the correlation parameter $\rho$ is bounded between [-1,1]. The determinant for this matrix can be obtained as \cite{ferdowsi2019rho}
\begin{equation}
\operatorname{det}\left(\mathbf{C}_{(\rho, s)}\right)=s^{d}\left(1-\rho^{2}\right)^{d-1},
\end{equation}
based on which we can derive the regularization term of the loss function as
\begin{equation}\label{lossNew}
\begin{aligned}
&\mathrm{KL}[q_{{\phi}}(\boldsymbol{z}_{l}\vert \boldsymbol{x}_{l})\Vert p_{\boldsymbol \theta}(\boldsymbol{z}_{l})]\!= \frac{1}{2}\left[d\left(-1-\log(\frac{s}{\boldsymbol{\sigma}_{prior}})+\right.\right.\\ &\left.\left.\frac{s^2}{\boldsymbol{\sigma}^2_{prior}}\right)+\frac{(\boldsymbol{\mu}_l\!-\!\boldsymbol{\mu}_{prior})^{2}}{\boldsymbol{\sigma}^2_{prior}}\!-(d-1) \!\log (1-\rho^{2})\right].
\end{aligned}
\end{equation}
Within the developed model for the MEC network, each BS collects a dataset including wireless transmission and computing delays from previous communications, and uses this dataset to train the proposed correlated VAE model. In order to evaluate the joint distribution of the transmission and computing latencies given a data point $\boldsymbol{x}$, $q_{{\phi}}(\boldsymbol{z}|\boldsymbol{x})$, we use the 1-dimensional convolutional network (CNN) encoder with two convolutional layers. Each convolution is followed by the rectified linear unit layer that introduces nonlinearity into the extracted features. These extracted features are concatenated into a single vector which is connected to a fully connected (FC) layer. Then, there is a decoder network, $p_{\boldsymbol{\theta}}(\boldsymbol{x}|\boldsymbol{z})$, that mirrors the encoder to reconstruct the input $\boldsymbol{x}$ back from the latent space sample $\boldsymbol{z}$. 

Using the proposed learning framework, the CDF of the E2E service delay $F_{\tau(m,n)}(.)$, defined in (\ref{R}), can be easily obtained from $q_{\phi}(\boldsymbol{z}|\boldsymbol{x})$, i.e., the joint distribution of the transmission and computing latencies. 

\begin{algorithm}[ht]
	\caption{Proposed Algorithm for Joint Task Assignment and Computing Resource Allocation}
	\label{alg:exmp}
	\renewcommand{\algorithmicrequire}{\textbf{Input:}}
	\renewcommand{\algorithmicensure}{\textbf{Output:}}
	\begin{algorithmic}[1]
		\REQUIRE $\mathcal{N}$, $\mathcal{M}$, $\boldsymbol{x}$ 
		\ENSURE $\boldsymbol{v}$ and $\boldsymbol{f}$ 
		\WHILE{$\boldsymbol{\theta}$ and $\phi$ are not converged }
		\STATE Randomly select minibatch of dataset $\boldsymbol{x}$.
		\STATE Feed the selected minibacth into CNN encoder $q_{{\phi}}(\boldsymbol{z}\vert \boldsymbol{x})$.
		\STATE Randomly sample $\epsilon \sim \mathcal{N}(\boldsymbol{\mu}_{prior},\boldsymbol{\sigma}_{prior}\mathbf{I_d})$.
		\STATE Calculate latent representation $\boldsymbol{z}$.
		\STATE Feed $\boldsymbol{z}$ into CNN decoder $p_{\boldsymbol{\theta}}(\boldsymbol{x}\vert \boldsymbol{z})$.
		\STATE Calculate VAE loss function in (\ref{lossNew}), update $\boldsymbol{\theta}$ and $\phi$.
	\ENDWHILE 
		\WHILE{maximum number of iterations is not reached}
			\STATE Using the distribution of E2E service delay obtained from Steps 1 to 8, solve the task assignment in (\ref{eq:3-OPT:a})-(\ref{eq:3-OPT:f}) and find $\boldsymbol{v^*_{mn}}$.
			\STATE Substitude $\boldsymbol{v^*_{mn}}$ into (\ref{eq:2-OPT:a}) and solve computing resource allocation in (\ref{eq:4-OPT:a})-(\ref{eq:4-OPT:b}).
		\ENDWHILE
	\end{algorithmic}
\end{algorithm}
\section{VAE-Based Reliability Optimization in MEC Networks}
Using the learned correlated VAE models, in this section, we propose a new optimization problem  based on risk theory to minimize the \emph{risk} as the performance metric, thus maximizing the reliability of MEC networks. 
\vspace{-0.15cm}
\subsection{Problem Formulation}
We build our problem formulation on the risk theory \cite{hao2021risk} that uses the concept of CVaR as a risk measure to characterize the tail distribution of the E2E service delay $\tau$. The CVaR provides the average of loss that exceeds the Value-at-Risk (VaR). The $\alpha$-VaR is the $\alpha$-percentile of distribution of a random variable given by \cite{rockafellar2002conditional}:\vspace{-0.1cm}
\begin{equation}\label{CVaR0}
\operatorname{VaR}_{\alpha}(\tau(m,n))\!=\!\arg \inf _{\tau_{\text{th}}}\{\tau_{\text{th}}: 1-\operatorname{F}_{\tau(m,n)}(\tau_{\text{th}}) \leq \alpha\},
\end{equation}
where $\alpha \in (0,1)$. In addtion, the function $\operatorname{F}_{\tau(m,n)}(\tau_{\text{th}})$, defined in (\ref{R}), is the CDF of the service delay that is learned by the VAE approach proposed in Section III. The CVaR function is defined as\vspace{-0.15cm}
\begin{equation}\label{CVaR1}\!\! 
\operatorname{CVaR}_{\alpha}(\tau(m,n))\! =\!\mathbbm{E}\left[\tau(m,n)\!\mid\!  \tau(m,n)\! >\!\operatorname{VaR}_{\alpha}(\tau(m,n))\right].\! 
\end{equation}
Next, we define an auxiliary function\vspace{-0.15cm}
\begin{equation}\!
\phi_{\alpha}(\tau(m,n), \tau_{\text{th}})\!:=\!\tau_{\text{th}}\!+\!\frac{1}{1-\alpha} \mathbbm{E}\left[(\tau(m,n)\!-\!\tau_{\text{th}})^{+}\right],\!\vspace{-0.15cm}
\end{equation}
where $(x)^{+}=\max (0, x)$  and the expectation is taken with respect to the distribution of the channel gain. According to \cite{rockafellar2002conditional}, the CVaR of $\tau(m,n)$ can finally be calculated as \vspace{-0.25cm}
\begin{equation}\label{CVaR2}
\operatorname{CVaR}_{\alpha}(\tau(m,n))=\min_{\tau_{\text{th}} \in \mathbbm{R}} (\phi_{\alpha}(\tau(m,n)), \tau_{\text{th}}).\vspace{-0.15cm}
\end{equation}
Meanwhile, we denote  $\boldsymbol{v}$ as an association vector defined as 
\begin{equation}
\boldsymbol{v}_{m n}=\left\{\begin{array}{ll}
1, & \text { If UE } m \text { is assigned to BS } n, \\
0, & \text { otherwise. }\vspace{-0.12cm}
\end{array}\right.
\end{equation}

To optimize the network reliability, our goal is to minimize the maximum risk among all the UEs, which can be described by the following optimization problem:
\begin{subequations}
 \begin{IEEEeqnarray}{rCl}\label{eq:2-OPT0}
\min_{\boldsymbol{v,f}}\max_{m \in \mathcal{M}}\,\,\, &&\boldsymbol{v}_{mn}\beta\text{CVaR}_\alpha(\tau(m,n))
\label{eq:2-OPT:a}\\
\text{s.t.,}\,\,\, 
&&\sum_{n \in \mathcal{N}} v_{m n} \leq 1, \forall m \in \mathcal{M}, \label{eq:2-OPT:b}\\
&& \sum_{m \in \mathcal{M}} v_{m n} \leq M, \forall n \in \mathcal{N},\label{eq:2-OPT:c}\\
&& \sum_{m \in \mathcal{M}} f(m,n) \leq f_{max}, \forall n \in \mathcal{N}, \label{eq:2-OPT:d}\\
&& v_{m n}\in\{0,1\} \label{eq:2-OPT:f},\vspace{-0.15cm}
\end{IEEEeqnarray}
\end{subequations}
where the objective function represents the risk measure for the MEC network and depends on the tail distribution of the service delay $\tau$ (derived via the correlated VAE) and $\beta \in (0,1)$ is the weight of the CVaR. Moreover,  (\ref{eq:2-OPT:b}) and (\ref{eq:2-OPT:c}) represent, respectively, the constraint to ensure that each UE is associated to at most one BS, and each BS can serve up to $M$ UEs. (\ref{eq:2-OPT:d}) indicates that the sum of the  computation frequency allocated to UEs should not exceed the maximum computation frequency. 

The proposed optimization problem in (\ref{eq:2-OPT0})-(\ref{eq:2-OPT:f}) is a non-convex MINLP, hence, it is difficult to solve. Next, we develop a new efficient algorithm to solve this problem.
\vspace{-0.05cm}
\subsection{Proposed Algorithm for Joint Task Assignment and Resource Allocation}
The proposed algorithm is summarized in Algorithm 1 which decomposes the problem into two sub-problems that solve the task and computing resource allocation problems within MEC. The first sub-problem is formulated as
\begin{subequations}
\begin{IEEEeqnarray}{rCl}\label{eq:3-OPT0}
\min _{\boldsymbol{v}} \max_{m \in \mathcal{M}} & \sum_{n \in \mathcal{N}} {\boldsymbol{v}_{m n}} \beta\text{CVaR}_\alpha(\tau_t(m,n))
\label{eq:3-OPT:a}\\
\text { s.t. } & \sum_{n \in \mathcal{N}} v_{m n} \leq 1, \forall m \in \mathcal{M}, \label{eq:3-OPT:b}\\
&\sum_{m \in \mathcal{M}} v_{m n} \leq M, \forall n \in \mathcal{N},\label{eq:3-OPT:c}\\
& 0 \le v_{m n}\le 1 \label{eq:3-OPT:f}.
\end{IEEEeqnarray}
\end{subequations}
The constraint (\ref{eq:3-OPT:f}) is obtained from the integer programming relaxation of the constraint (\ref{eq:2-OPT:f}). The optimization problem in (\ref{eq:3-OPT0})-(\ref{eq:3-OPT:f}) can be solved via applying the Karush-Kuhn-Tucker (KKT) conditions. After finding the optimal association vector, $\boldsymbol{v}^*_{mn}$, we substitute it into (\ref{eq:2-OPT0}) and formulate the second sub-problem as follows:
\begin{subequations}
\begin{IEEEeqnarray}{rCl}\label{eq:4-OPT0}
\min _{\boldsymbol{f}} \max_{m \in \mathcal{M}} &\sum_{n \in \mathcal{N}} \boldsymbol{v}^*_{mn}\beta\text{CVaR}_\alpha(\tau_p(m,n))
\label{eq:4-OPT:a}\\
\text { s.t. } & \sum_{m \in \mathcal{M}} f(m,n) \leq f_{max}, \forall n \in \mathcal{N}. \label{eq:4-OPT:b}\vspace{-0.2cm}
\end{IEEEeqnarray}
\end{subequations}
Since $\tau_{p}(m, n) = \frac{c_m}{f(m, n)}$, the problem in (\ref{eq:4-OPT0}) is a non-convex optimization problem. To transform it into a convex problem, we use an auxiliary variable $g(m,n)=[\frac{1}{f(m,n)}]$, $\forall m \in \mathcal{M}$ and  $\forall n \in \mathcal{N}$. Then, the computing resource allocation problem can be reformulated as
\begin{subequations}
\begin{IEEEeqnarray}{rCl}\label{eq:5-OPT0}
\min _{{g}} \max_{m \in \mathcal{M}} &\sum_{n \in \mathcal{N}} \boldsymbol{v}^*_{mn}\beta\text{CVaR}_\alpha(c_m{g}(m,n))
\label{eq:5-OPT:a}\\
\text { s.t. } & \sum_{m \in \mathcal{M}} \frac{\boldsymbol{v}^*_{mn}}{g(m,n)} \leq f_{max}, \forall n \in \mathcal{N}, \label{eq:5-OPT:b}
\end{IEEEeqnarray}
\end{subequations}
which is a convex optimization problem.
\vspace{-0.2cm}
\section{Simulation Results}
In this section, we evaluate the performance of the proposed scheme for reliability optimization in MEC networks. As a case study, we consider a network of CAVs that leverages an MEC network to manage processing tasks required for their autonomous navigation. The
downlink and uplink packet sizes are selected randomly from a uniform distribution with a range $[1,10]$ kbits. Simulation parameters are summarized in Table \ref{tabsim1}. We compare the performance of the proposed method with two baseline approaches. The first baseline approach, hereinafter referred to as ``Baseline 1'', aims to minimize the maximum average service delay among all the CAVs, , i.e., $\mathbbm{E}\left[\tau(m,n)\right]$. The second baseline, hereinafter referred to as ``Baseline 2'', aims to minimize an upper bound (derived in \cite{hao2021risk}) for the weighted sum of the average service delay
and the CVaR, i.e., $\mathbbm{E}\left[\tau(m,n)\right]+\beta \mathrm{CVaR}_{\alpha}\left(\tau(m,n)\right)$. A detailed description of the two baseline approaches is available in \cite{hao2021risk}.

Therefore, the baseline methods rely only on the average performance metrics and do not take into account the VAEs' outcomes for solving the joint task assignment and computing resource allocation. The performance was evaluated by averaging the results over sufﬁciently large Monte Carlo runs. 
\begin{table}[t!]
	\footnotesize
	\centering
	\caption{\vspace*{-0cm}  Simulation Parameters}\vspace*{-0.2cm}
	\begin{tabular}{|>{\centering\arraybackslash}m{0.9cm}|>{\centering\arraybackslash}m{3.7cm}|>{\centering\arraybackslash}m{2.5cm}|}
		\hline
		\bf{Notation} & \bf{Parameter} & \bf{Value} \\
		\hline
		$N$    & Number of BSs &  10\\
		\hline
		$M$ & Number of CAVs &  40 \\
	    \hline
		$P_n$    & Transmit power of a BS& 100 mW \\
		\hline
		$P_m$    & Transmit power of a CAV& 10 mW\\
		\hline
		$G_n$,$G_m$      & Antenna gains & 1\\
		\hline
		$f_{max}$   & Computing frequency & 20 GHz \\
		\hline
		$N_0$ &  Noise power spectral density & $-90$ dBm/Hz \\
		\hline
		$W$ &  Total system bandwidth & 100 MHz \\
		\hline
		$\tau_{\text{th}}$ & Service delay requirement& 10-100 ms\cite{semiari2019integrated}\\
		\hline
	\end{tabular}\label{tabsim1}\vspace{-0.35cm}
\end{table}
\subsection{Training data}
In order to compute the computational latency in a realistic scenario, we use one of the most powerful object detection CNN algorithms, namely YOLOv3 (You Only Look Once). Here, we consider object detection since it is one of the common processing tasks in applications such as CAVs or XR that rely on computer vision. 
 
We use Berkeley Deep Drive 100k (BDD100K) dataset to train the CNN. BDD100K consists of more than 100,000 images of size 1280×720 with ten different classes (bus, traffic light, traffic sign, person, bike, truck, motorcycle, car, train, rider) with a training set of 70,000 image-label pair, a test set of 20,000 image-label pair and a validation set of 10,000 data. We use ImageNet pre-trained network parameters given in repository and train the model over 10 epochs using a deep learning server equipped with i9-9920X X-series CPU, 128GB DDR4 memory, and two Quadro RTX 6000 GPUs plus NVLink.
\subsection{Results and discussion}
\begin{figure}[t!]
	\centering
	\centerline{\includegraphics[width=8cm]{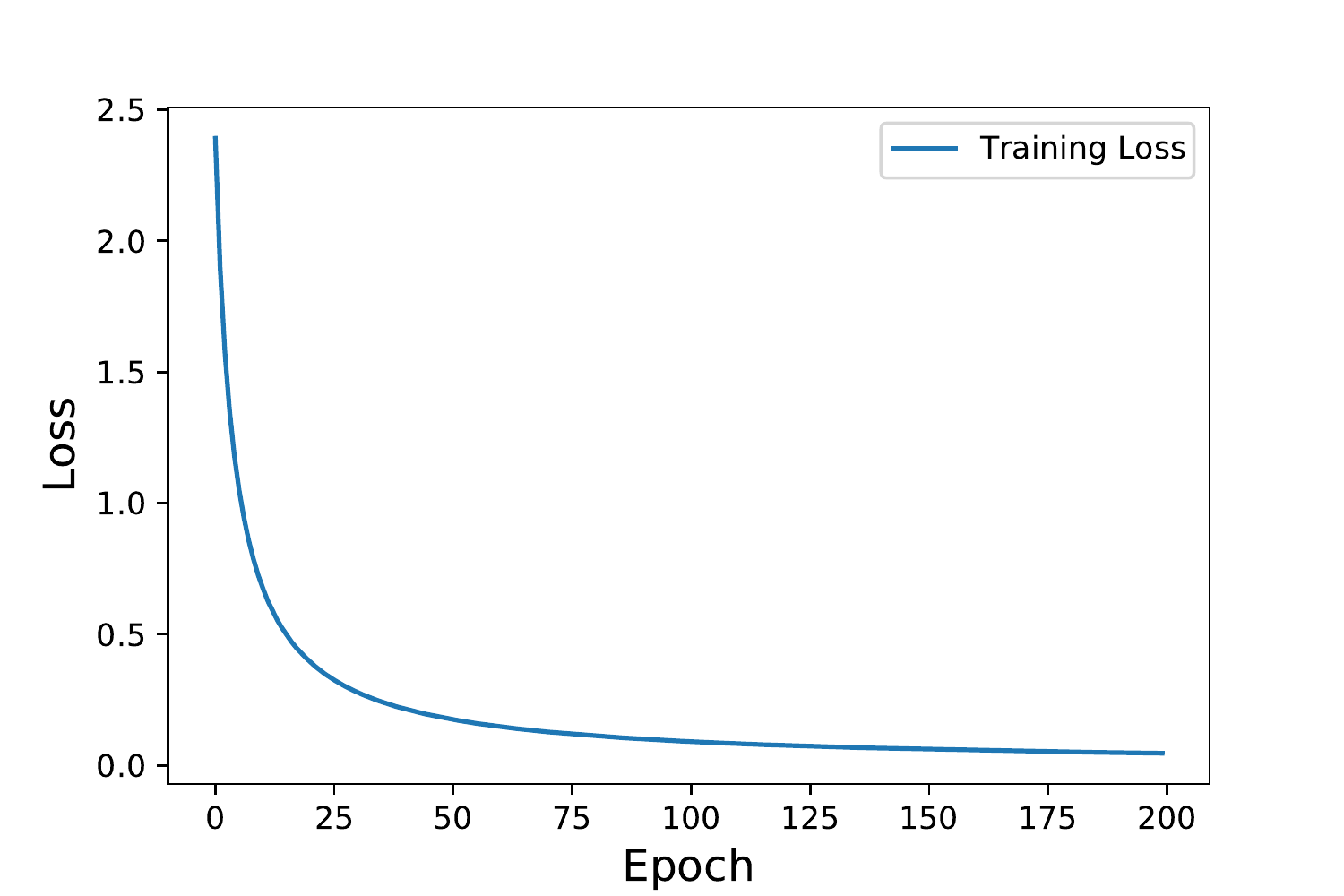}}\vspace{-0.5em}
	\caption{\small Correlated VAE loss function versus the number of epochs.}\vspace{-1.2em}
	\label{VAELOSS}
\end{figure}

Figure. \ref{VAELOSS} shows the average loss function for the training of the correlated VAE versus the number of epochs. Here, the average loss is computed by averaging $\mathcal{L}$ over large independent runs. From Fig. \ref{VAELOSS}, we observe that the loss decreases rapidly, showing the fast convergence of the VAE. The results show that the proposed correlated VAE successfully converges within reasonably small number of epochs.   

\begin{figure}[t!]
	\centering
	\centerline{\includegraphics[width=7cm,height=5cm]{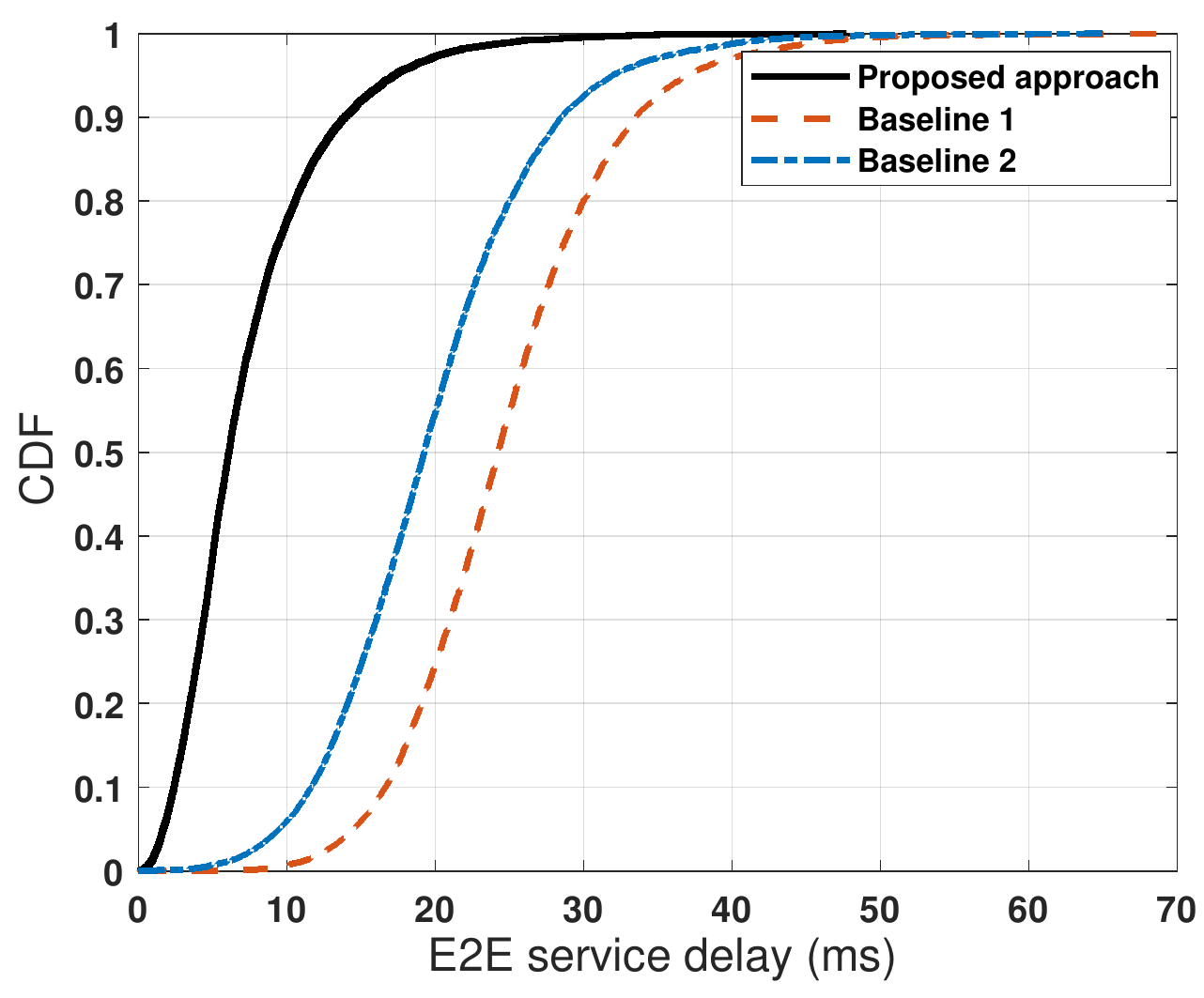}}\vspace{-0.6em}
	\caption{\small CDF of the E2E service delay.}\vspace{-1em}
	\label{CDF}
\end{figure}

\begin{figure}[t!]
	\centering
	\centerline{\includegraphics[width=7cm,height=5cm]{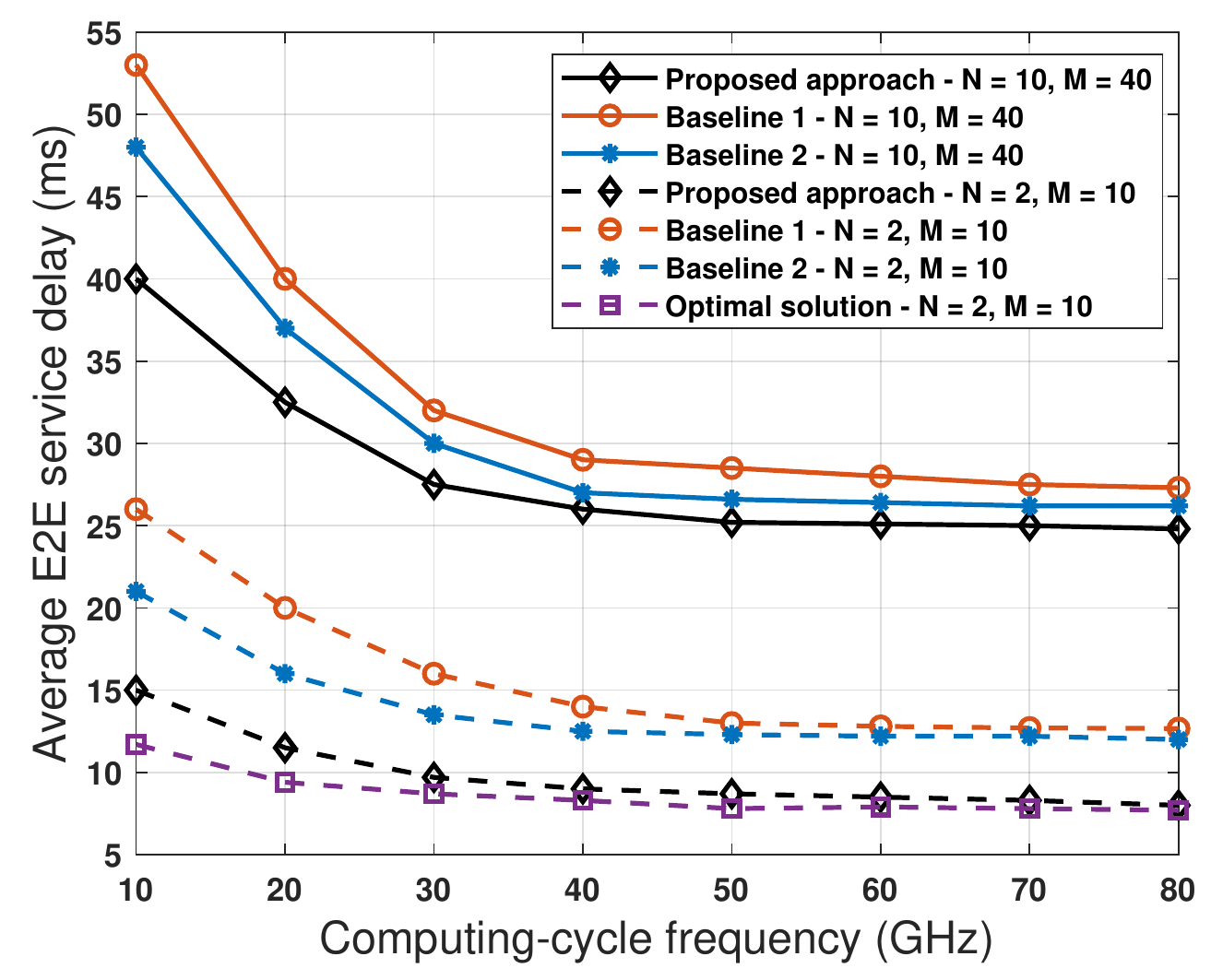}}\vspace{-1em}
	\caption{\small Average E2E service delay versus the computing frequency.}\vspace{-1em}
	\label{COMFREQ}
\end{figure}

Figure. \ref{CDF} compares the CDF of the E2E service delay, and equivalently, the network's reliability, resulting from the proposed approach and the two baseline methods. The results in Fig. \ref{CDF} show that the proposed approach completely outperforms the baseline schemes. For example, we can observe that the proposed approach can guarantee 30 ms E2E latency with a probability of 0.9999, ensuring a high reliability for the MEC network. However, both baseline approaches can only satisfy this E2E latency requirement with probabilities less than 0.92. In addition, as shown in Fig. \ref{CDF}, baseline 2 performs better compared to baseline 1 indicating that adding the CVaR to the optimization objective can reduce the risk of high delays.

Figure. \ref{COMFREQ} compares the average E2E service delay for the proposed approach with the baseline methods, versus the computing-cycle frequency of each BS. It can be seen that by increasing the computing-cycle frequency, the E2E service delay decreases for all the three methods. Furthermore, we note that the E2E service delay does not reach zero when the computing-cycle frequency is relatively high. This is because for a high computing-cycle frequency, the total delay is mainly determined by a non-zero transmission delay. The results in Fig. \ref{COMFREQ} also show that the proposed algorithm outperforms the other two schemes. For example, for $f$=30 GHz, the performance gain is up to 18\% and 10\%, respectively, compared to baseline methods 1 and 2 when $N=10$ BSs and $M=40$ CAVs.  More importantly, the results verify the near-optimal performance of the proposed method compared with the optimal solution obtained from an exhaustive search for a network with $N=2$ BSs and $M=10$ CAVs. For instance, the optimality gap for $f=70$ GHz is only $5\%$ and it reduces as the computing cycle increases. Also, the performance gains compared to baseline methods 1 and 2 are $35\%$ and $30\%$, respectively.

\begin{figure}[t!]
	\centering
	\centerline{\includegraphics[width=7cm,height=5cm]{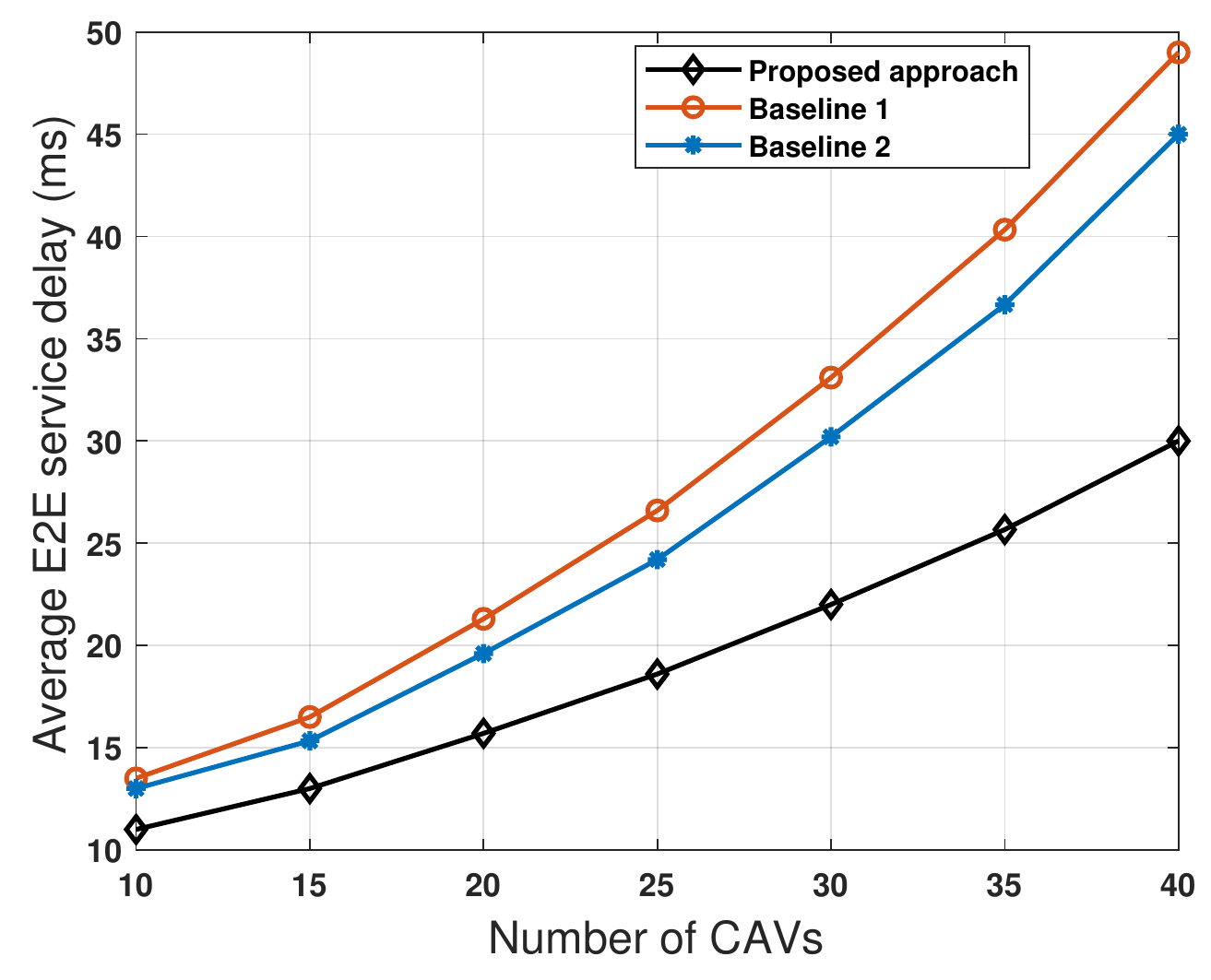}}\vspace{-1em}
	\caption{\small Average E2E service delay versus the number of CAVs.}\vspace{-1.5em}
	\label{Hist}
\end{figure}
\begin{figure}[t!]
	\centering
	\centerline{\includegraphics[width=8cm]{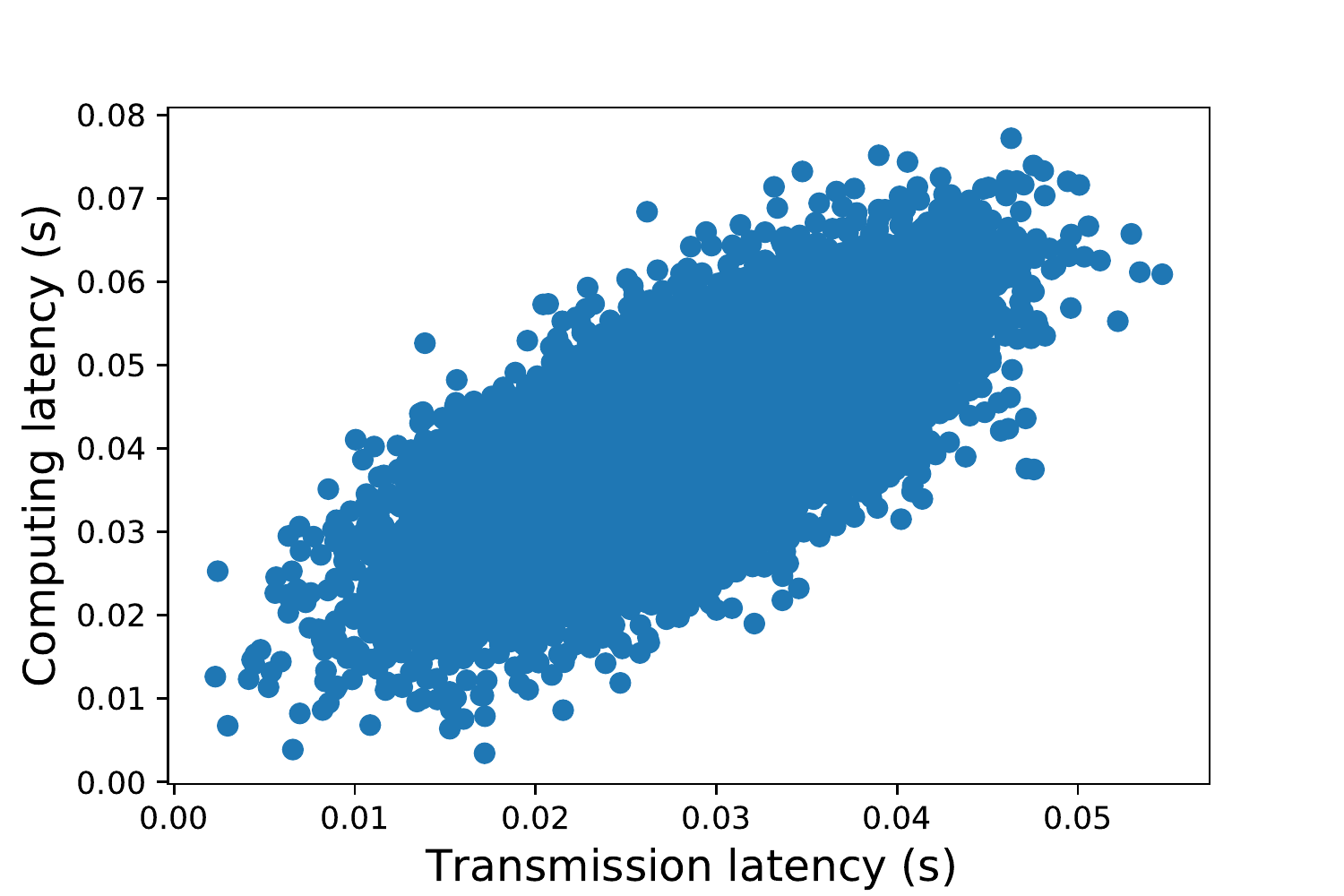}}\vspace{-1em}
	\caption{\small Scatter plot of the mean for the transmission and computing latencies within the correlated VAE's latent space.}\vspace{-1.5em}
	\label{Scatter}
\end{figure}

In Fig. \ref{Hist}, the average E2E service delay per CAV versus the network size is shown for the proposed approach and the two baseline methods. Clearly, the average E2E service delay 
increases as more CAVs exist in the network. The results in Fig. \ref{Hist} show the superior performance of the proposed approach compared to the baseline methods. For example, in a network with $M=30$ CAVs, the performance gains yielded by the proposed algorithm are up to 35\% and 27\%, respectively, compared to baseline schemes 1 and 2. Furthermore, Fig. \ref{Hist} also highlights the scalability of the proposed scheme. For example, with the average E2E service delay threshold of $30$ ms, the proposed algorithm can support up to $40$ CAVs, which is $40\%$ and $32\%$ higher compared to baselines 1 and 2, respectively.  

Figure. \ref{Scatter} shows the scatter plot of the expected values of latencies within the 2-dimensional latent space, obtained from the correlated VAE. The x-axis and y-axis represent, respectively, the distribution of the transmission delay and the computing delay. Here, each point represents the outcome from one learning epoch. From Fig. \ref{Scatter}, we can observe that the proposed correlated VAE can successfully capture the correlation between input data points for the transmission and computing latencies, resulting from unbalanced task assignment within the MEC network. We can also observe that, the derived correlation between two latent variables (i.e., transmission and computing latencies) is positive.
\section{Conclusions}
In this paper, we have developed a novel framework to optimize the E2E reliability in MEC networks. The proposed scheme has adopted a learning method, based on correlated VAEs, to estimate the distribution of the E2E service delay. Then, a new problem has been formulated that uses the trained VAE models to maximize the network reliability by minimizing the CVaR as a risk measure of the E2E service delay. To solve the proposed non-convex MINLP, we have developed a new algorithm that iteratively solves the task allocation problem within the MEC network jointly with computing resource allocation at each edge computing server. Simulation results have confirmed the effectiveness of the developed VAE method for estimating the distribution of the E2E service delay and capturing the underlying correlations between the transmission and computing latencies. Furthermore, the results have shown that the proposed algorithm can substantially improve the network's  reliability compared to other baseline methods that ignore the distribution of the E2E service delay. \vspace{0cm}

\bibliographystyle{IEEEbib}
\bibliography{references}
\end{document}